\documentclass[pdftex,rmp,twocolumn,groupedaddress,floatfix]{revtex4}
\usepackage[utf8]{inputenc}
\usepackage[T1]{fontenc}
\usepackage{graphicx}
\usepackage{amssymb}

\usepackage{mathptm}
\usepackage{times}
\begin{document}

\title{The transmission sense of information}

\author{Carl T. Bergstrom}
\email{cbergst@u.washington.edu}
\altaffiliation[Also at ]{Santa Fe Institute, 1399 Hyde Park Rd., Santa Fe, NM
87501}
\homepage{http://octavia.zoology.washington.edu/}
\author{Martin Rosvall}
\email{rosvall@u.washington.edu}
\affiliation{Department of Biology, University of Washington, Seattle, WA
98195-1800}
\date{\today}
%\pacs{89.75.-k, 89.75.Fb, 89.70.+c}
\begin{abstract}

Biologists rely heavily on the language of information, coding, and transmission that is commonplace in the field of information theory as developed by Claude Shannon, but there is open debate about whether such language is anything more than facile metaphor. Philosophers of biology have argued that when biologists talk about information in genes and in evolution, they are not talking about the sort of information that Shannon's theory addresses. First, philosophers have suggested that Shannon theory is only useful for developing a shallow notion of correlation, the so-called ``causal sense'' of information. Second they typically argue that in genetics and evolutionary biology, information language is used in a ``semantic sense,'' whereas semantics are deliberately omitted from Shannon theory. Neither critique is well-founded. Here we propose an alternative to the causal and semantic senses of information: a {\em transmission sense of information}, in which 
an object X conveys information if the function of X is to reduce, by virtue of its sequence properties,  uncertainty on the part of an agent who observes X. The transmission sense not only captures much of what biologists intend when they talk about information in genes, but also brings Shannon's theory back to the fore. By taking the viewpoint of a communications engineer and focusing on the decision problem of how information is to be packaged for transport, this approach resolves several problems that have plagued the information concept in biology, and highlights a number of important features of the way that information is encoded, stored, and transmitted as genetic sequence.

\end{abstract}

\maketitle

\section{Introduction}

Biologists think in terms of information at every level of investigation. Signaling pathways transduce information, cells process information, animal signals convey information. Information
flows in ecosystems, information is encoded in the DNA, information is carried by nerve impulses. In some domains the utility of the information concept goes unchallenged: when a brain scientist
says that nerves transmit information, nobody balks. But when geneticists or evolutionary biologists use information language in their day-to-day work, a few biologists and many philosophers become anxious about whether this language can be justified as anything more than facile metaphor \citep{sterelny_sex_1999,Sterelny00,godfrey-smith_theoretical_2000,griffiths_2001,griesemer2005,godfrey-smith_hull_cambridge_2008}. Why do the neurobiologists get a free pass while evolutionary geneticists get called on the carpet? When neurobiologists talk about information, they have two things going for them. First, there is a straightforward analogy between electrical impulses in neural systems and the classic communications theory picture of source, channel, and receiver \citep{shannon1948}. Second, information theory has obvious ``legs'' in neurobiology: for decades neurobiologists have profitably used the theoretical apparatus of information theory to understand their study systems. Geneticists are not so fortunate. For them, the analogy to communication theory is less obvious. Efforts to make this analogy explicit seem forced at best  and the most successful uses of information-theoretic reasoning within the field of genetics rarely make explicit their information-theoretic foundations or make use of information-theoretic language \cite{CrickEtAl57,Kimura61,Felsenstein71,freeland_genetic_1998}. 

As a consequence, philosophers have concluded that the mathematical theory of communication pioneered by Claude Shannon in 1948 (hereafter ``Shannon theory'') is inadequate to ground the notion of information in genetics and evolutionary biology. First, philosophers have unfairly suggested that Shannon theory is only useful for developing a shallow notion of correlation, the so-called ``causal sense'' of information. Second, they typically argue that in genetics and evolutionary biology, information language is used in a ``semantic sense'' --- and of course semantics are deliberately omitted from Shannon theory. 

Neither critique is well-founded. In this paper we begin by summarizing the causal and semantic views of information. We then propose an alternative --- a transmission sense of information --- that not only captures much of what biologists intend when they talk about information in genes, but also brings Shannon's theory back to the fore.

\subsection*{Causal view of information}

Inspired by Dretske \cite{Dretske83}, several authors \cite{GriffithsAndGray94,sterelny_sex_1999,griffiths_2001,godfrey-smith_hull_cambridge_2008} have explored  Shannon theory as a grounding for information language in biology. They derive roughly the following picture: The key currency in information theory is entropy $H(X)$ and the key statistic is mutual information $I(X;Y)$, where $X$ and $Y$ are random variables. Information is conveyed in Grice's sense of natural meaning \cite{Grice57}: whenever $Y$ is correlated with $X$, we can say that $Y$ carries information about $X$. There is no deep notion of meaning or coding here. ``[W]hen a biologist introduces information in this sense to a description of gene action or other processes, she is not introducing some new and special {\em kind} of relation or property'', Godfrey Smith writes, ``She is just adopting a particular quantitative framework for describing ordinary correlations.''  \cite{godfrey-smith_hull_cambridge_2008}. In this {\em causal sense} of information,  genes carry information about phenotypes just as smoke carries information about fire, nothing more. If this is all that biologists are doing when they talk about how the genotype carries information about phenotype, this is a shallow use of the information concept!

\begin{figure}[hptb]
\centering
\includegraphics[width=0.7\columnwidth]{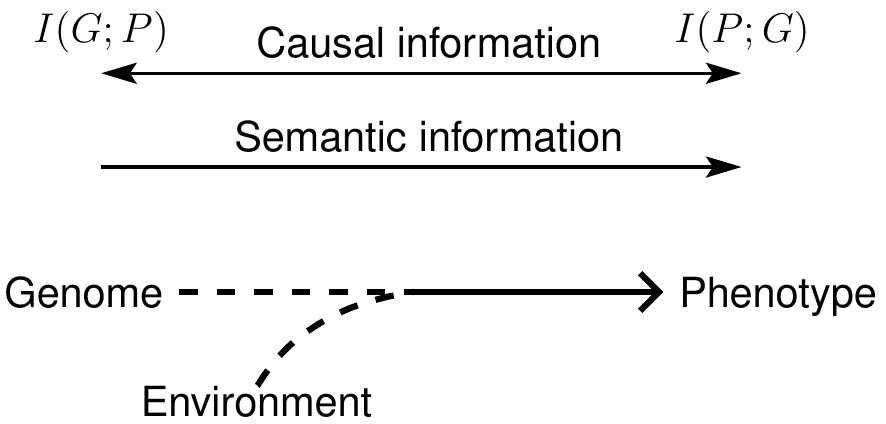}
\caption{\label{fig1}\itshape Information theory restricted to a descriptive statistics for
correlations. Information flows in both directions between genotype and phenotype, $I(P;G) = I(G;P)$,
and, according to the parity thesis, there is nothing that privileges genes over environment.
}
\end{figure}

Not only is this sense shallow, it fails to capture the directional flow of information from genotype to phenotype that is the central dogma of molecular biology \cite{Crick70}. If by ``G has information about P'' we mean only that $I(G;P)>0$, then we are not acknowleding the direction of information from genotype to phenotype 
\citep{griffiths_2001,godfrey-smith_theoretical_2000,godfrey-smith_hull_cambridge_2008}. The reason is that the mutual information $I$ is by definition a symmetric quantity; $I(G;P)=I(P;G)$. Mathematically, the amount of information that knowing the genotype $G$ provides about the phenotype $P$ is always exactly equal to the amount of information that knowing the phenotype $P$ provides 
about the genotype $G$ (see Fig.~\ref{fig1}).

Worse still, this notion of mutual information fails to capture the sense of privilege that biologists tend to ascribe to the informational molecule DNA over other contributors to phenotype. So far as causal covariance is concerned, both genes $G$ and environment $E$ influence phenotype $P$ --- and in principle we can equally well compute either $I(G;P)$ and $I(E;P)$. So it seems that Shannon theory has no way of singling out DNA as an information-bearing entity. 

This criticism is formalized as the parity thesis, and is crafted around an important result in information theory that the roles of source and channel conditions are exchangeable \citep{GriffithsAndGray94}. Typically when one sits in front of the television, the football broadcast is the signal. The weather, a crow landing on the television antenna, interference from a neighbor's microwave --- these are sources of noise, the channel conditions for our transmission. But a television repairman has an opposite view. He doesn't want to watch the game, he wants to learn about what is altering the transmission from the station. So he tunes  your set to a station broadcasting a test pattern. For the repairman, this test pattern provides channel conditions to read off the signal of how the transmission is being altered. As Sterelny and Griffiths point out (1999), ``The sender/channel distinction is a fact about our interests, not a fact about the physical world.'' The parity thesis applies this logic to genes and environment. In the parity view, whether it is genome or environment that carries information must be a fact about our interests, not a fact about the world. 

The problem with these arguments is that they adopt a few tools from Shannon theory, but neglect its {\em raison d'\^etre}: the underlying decision problem of how to package information for transport.  Before delving deeper into Shannon theory, we will take a brief detour to summarize the semantic sense of information in biology.

\subsection*{Semantic view of information}

In addition to the limitations enumerated above, the causal view fails to highlight the intentional, representational nature of genes and other biological objects \cite{GriffithsAndGray94,sterelny_sex_1999,Shea07,godfrey-smith_hull_cambridge_2008}. When biologists talk about genes as informational molecules, this argument goes, it is not because they are correlated with other things (e.g. amino acid sequence or phenotype), but rather because they {\em represent} other things. This semantic sense of information in which ``genes semantically specify their normal products'' \cite{GodfreySmith07} cannot be captured using Shannon theory, which is by design silent on semantic matters.

But what is it that genes are supposed to represent? Much of the conventional language of molecular biology suggests phenotypes as an obvious candidate, and this is the approach that Maynard Smith takes in a target article that triggered much of the recent debate over the information concept in biology \cite{maynardsmith2000}.  At first glance, this view has several things to recommend it. A semantic notion of genetic information captures the directionality discussed above: genes represent
phenotypes but phenotypes do not represent genes. One could also try to argue that the semantic view privileges genes in that we can say that genes have a representational message about phenotype but environment does not. Finally, it allows for misrepresentation or false representation, whereas causal information does not \cite{griffiths_2001}\footnote{We think that Stegmann handles the misrepresentation issue even more cleanly with a shallow semantic notion of genes as conveying instructional, as opposed to representational, content \cite{stegmann04}}. 

Does this mean that the problem is solved? No --- \citet{griffiths_2001} and
\citet{godfrey-smith_hull_cambridge_2008} argue that semantic information remains vulnerable to the parity thesis. 
Moreover, Godfrey-Smith (1999,2008) and Griffiths (2001) note that the reach of the semantic information concept within genetics is very shallow: legitimate talk of semantic representation can go no further than genes coding for
amino acids\nocite{GodfreySmith99,griffiths_2001,godfrey-smith_hull_cambridge_2008}. Beyond this point, the mapping from genotype forward is context-dependent and hopelessly entangled in a mass of causal interactions. Thus, these authors conclude, the relation from genes to phenotype cannot be a representational one.  Accordingly, it seems as if the semantic view of information has been pushed as far as it will go, and yet we are left without a fully satisfactory account of the information concept in biology. Let us therefore return to Shannon's information theory, but move beyond the causal sense.

\section{A transmission view of information}

As we described above, philosophers of biology largely restrict Shannon theory to a descriptive statistics for correlations.
This misses the point. At the core of Shannon theory is the study of how far
mathematical objects such as sequences and functions can be compressed without
losing their identity, and if compressed further, how much their structure will be distorted.
From this foundation in the limits of compression emerges a richly
practical theory of coding: information theory is a decision theory of how to
package information for transport, efficiently.  It is a theory about the
structure of those sequences that efficiently transmit information. And it is a
theory about the fundamental limits with which that information can be transmitted \cite{ShannonAndWeaver49,Pierce80,Yeung02,CoverAndThomas06}.

This decision-theoretic view of Shannon theory is missing from the
discussions of information in biology.  We will argue that this view
provides a justification for information language as applied to
genes, and that it also resolves the apparent and unappealing
symmetries of (1) mutual information between genes and phenotype and
(2) the parity thesis. 

In the original formulation of Shannon theory, information is what an agent packages for transmission through a channel to reduce uncertainty on the part of a receiver.  This information is physically instantiated and spatiotemporally bounded. Thus, as Lloyd and Penfield (2003) note\nocite{Lloyd03}, information can be sent either from one {\em place} to another, or from one {\em time} to another\footnote{In practice, it  takes time to send information from one place to another, but the conventional Shannon framework suppresses this time dimension.}. Usually when we talk about sending information from one place to another, we posit two separate actors, one of whom sends a message that the other receives; when we talk about sending information from one time to another, we posit a single agent who stores information that she herself can later retrieve. But whether the message goes across space or time, whether there are one or two agents involved, whether we use the language of signal transmission or the language of data storage, mathematically these are exactly the same process.  Thus in practice, when you package information and then send it either across the space dimension as a signal or across the time dimension via storage and retrieval,  you are  {\em transmitting} information.

Think about what happens when you send a message to your friend by burning a compact disc. Your computer encodes a message onto the digital medium. You send the medium through a channel (e.g. the postal service). Your friend, the receiver, puts the CD into her computer, the computer decodes the message, and she hears the sweet strains of Rick Astley.  But it doesn't matter that you sent the disc through the mail -- all of the mathematical operations that underly the information encoding are the same whether you send the CD to a friend or save it for your own later use. To cross space or time, we can encode the same way. Indeed, we use the same error-correcting codes for storage and retrieval on CDs  as we do for sending digital images from deep space back to Earth \cite{Cipra93}.

Taking this view of information and transmission, let us return to the proposed schematic of biological information in Figure 1. This picture has neither a space dimension or a time dimension; information is not being sent anywhere. Here we simply have a correlation (if one takes a causal view) or a translation (if one takes a semantic view). Thus within the actual Shannon framework, Figure 1 simply illustrates a decoder.
Likewise, notice that the biological processes underlying the schematic in Figure 1 are not the processes that biologists refer to when they talk about transmission. In biology, {\em transmission genetics} is the study of inheritance systems, not the study of transcription and translation, and {\em genetic transmission} is the passing of genes from one generation to another, not the passing of information from genotype to phenotype. 

\begin{figure}[hptb]
\centering
\includegraphics[width=0.8\columnwidth]{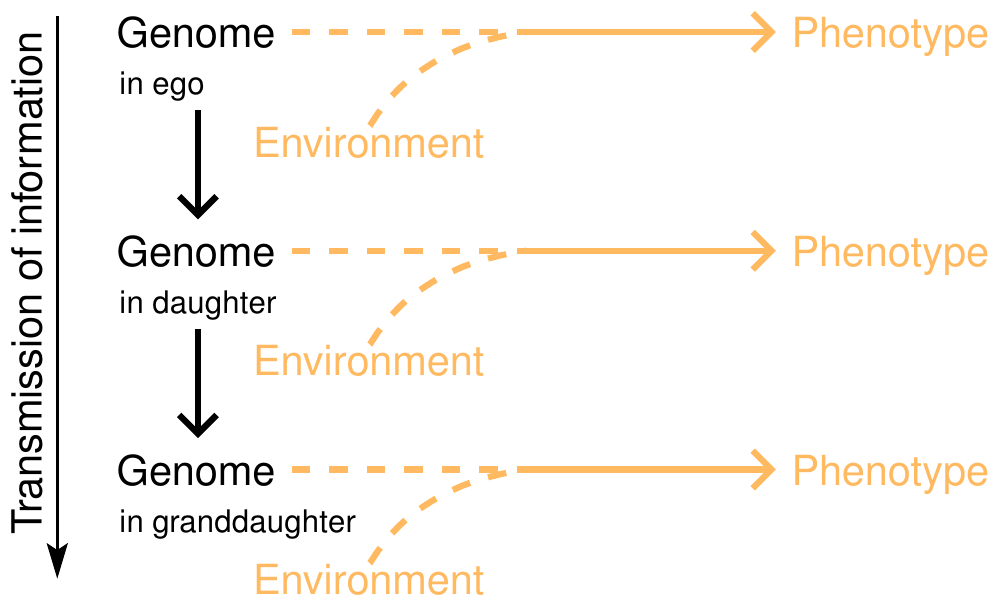}
\caption{\label{fig2}\itshape In biology, genetic transmission occurs vertically (from parent to offspring to grandoffspring). It is upon this axis that the transmission sense of information focuses. }
\end{figure}

In life and in evolution the transmission of information goes from generation to generation to
generation as in Figure \ref{fig2}. Here is the
transmission; we know that genes are transmitted from parent to
offspring in order to provide the offspring with
information about how to make a living (e.g.\ metabolize sugars,
create cell walls, etc.) in the world. This suggests that can make sense of a large fraction of the use of information language in biology if we adopt a {\em transmission view of information\footnote{A message need not be composed of multiple characters  to meet this definition. Even a string of length one is a sequence, thus even a single character conveys information.}}.

\bigskip
\begin{center}
\begin{minipage}{0.8\columnwidth}
{\bf Transmission view of information} \\
An object X conveys information if the function of X is to reduce, by virtue of its sequence properties,  uncertainty on the part of an agent who observes X.
\end{minipage}
\end{center}
\bigskip

As with many aspects of science, the tools and language that we use have a strong influence on the questions that we think to ask --- and once we shift to the transmission sense of information, our focus changes. When we view biological information as a semantic relationship, we are drawn to think like developmental biologists, about how information goes from an encoded form in the genotype to its expression in the phenotype. But when we talk about the transmission sense of information, we step out in an orthogonal direction and we can now see information as it flows through the process of intergenerational genetic transmission (Figure \ref{fig2}). And once we do that, we can start to think about natural selection, the evolutionary process, and how information gets into the genome in the first place.

\subsection*{Symmetry of mutual information}

By viewing Shannon information as a result of a decision process instead of as a
correlation measure, we can resolve the concern that Godfrey-Smith raises about the bidirectional flow of information in Shannon theory. Godfrey-Smith dismisses the causal sense of information because ``information in the Shannon sense `flows' in both directions, as it involves no more than learning about the state of one variable by attending to another'' \cite{godfrey-smith_hull_cambridge_2008}. Here Godfrey-Smith is referring to the symmetry of the mutual information measure: $I(X;Y)=I(Y;X)$ (Figure 1). 

What is the mutual information actually measuring when we apply this equality in a communication context?  An example helps. Peter and Paul are concerned about the state of the world $W$. Suppose that Peter observes a correlate $X$ of the random variable $W$. We want him to communicate his observation to Paul, and he does so using a signal $Y$. The mutual information $I(X;Y)$  tells us how effectively he conveys what he sees, {\em on average}, given the statistical distribution of possible $X$ values, the properties of the channel across which the signal is sent, etc. Specifically, $I(X;Y)=H(X)-H(X|Y)$ measures how much Paul learns by knowing $Y$ about what Peter saw, $X$, again on average. Because $I(X;Y)=H(X)-H(X|Y)=H(X)+H(Y)-H(X,Y)=H(Y)-H(Y|X)=I(Y;X)$, Peter knows exactly as much about what Paul learns as Paul learns about what Peter saw.  But $I(Y;X)$ is usually irrelevant when we think about the decision problem of communicating. In this context we want Peter to get a message about the world to Paul, and we rarely care how much Peter knows afterwards about what Paul has learned.

\begin{figure}[hptb]
\centering
\includegraphics[width=0.6\columnwidth]{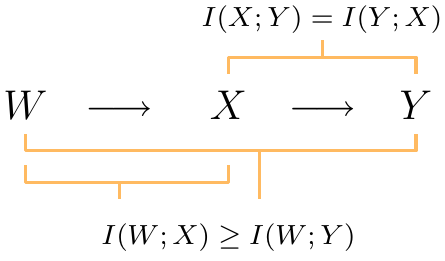}
\caption{\label{fig:dpi}\itshape Despite the symmetry of the mutual information $I(X;Y)=I(Y;X)$, the data processing inequality reveals the directional flow of information in Shannon's scheme.}
\end{figure}

This directionality is manifested within Shannon theory by the {\em data processing inequality} (Figure \ref{fig:dpi}) \cite{CoverAndThomas06,Yeung02}. This theorem states that the act of processing data, deterministically or stochastically, can never generate additional information.  A corollary pertains to communication: along a communication chain, information about the original source can be lost but never gained. In the scenario described above, both the observation step $W\rightarrow X$ and the communication step $X\rightarrow Y$ are steps in a Markov chain. For any Markov chain $W \rightarrow X \rightarrow Y$, the data processing inequality states that $I(W;X) \geq I(W;Y)$. In our example, the data processing inequality reveals that communication between Peter and Paul is not symmetric. Paul may know as much about what Peter sent as Peter knows about what Paul received, but Paul does not in general know as much about what matters --- the state of the world --- as Peter does. Shannon theory is not symmetric with respect to the direction of communication. 

\subsection*{The Parity thesis}

In the introduction to this paper, we described the parity thesis. While there is a good case to be made for parity between genes and environment when we restrict our view to the horizontal development of phenotype from genotype, that parity is shattered when we look along the vertical axis of intergenerational transmission.

Look at Figure 2, which corresponds to a neo-Darwinian view of evolution. In this model of the biological world, the transmission concept cleanly separates genes from environments. The former are transmitted across generations, the latter are not. Moreover, taking a teleofunctional view as \citet{SterelnyEtAl96} do for replicators in general, the hypothesis that genes are {\em for} transmission across generations is richly supported by the physical structure of the DNA. Genes are
made out of DNA, a molecule that is exquisitely fashioned so as to
(1) encode lots of sequence information in small space, (2) be
incredibly easy to replicate, (3) be arbitrarily and infinitely
extensible in what it can say, and (4) be structurally very stable
and inert \cite{Lewontin92}. In fact, DNA is perhaps the most
impressive known substance with respect to (1) and (2). No machine
can look at a protein and run off a copy; DNA is exquisitely adapted
so that a relatively simple machine, the DNA polymerase, does this at
great speed and high fidelity. Think about how amazing it is that PCR works. It is as if you could throw a hard drive in a water bath with a
few enzymes and a few raw materials, run the temperature through a few cycles, and pull out millions of identical hard drives.  DNA practically screams ``I am for storage and transmission!''  

One might object that Figure 2 conveys an over-simplified view of the world. This is true. A more sophisticated view of the evolutionary process allows for additional channels of intergenerational transmission and information flow: environments can be constructed and inherited \cite{OdlingSmeeEtAl03}. Non-genetic biological structures such as membranes and centrosomes 
are inherited \cite{GriffithsAndKnight98} and can even be argued to carry some information \cite{griesemer2005}. Methylation provides an extensive layer of markup on top of nucleic acid sequence. Developmental switches actively transduce environmental information into epigenetically heritable forms \cite{griffiths_2001}. 

But such an objection misses our point. Our aim with the transmission sense of information is not to  single out uniquely the genes as having some special property which we deny to all other biological structures, but rather to identify those components of biological systems that have information storage, transmission, and retrieval as their primary function. Methylation tags are obvious members of this information-bearing class: they carry information across generations in the transmission sense and this appears to be their primary function. Extrinsic features of the environment such as ambient temperature are obviously not members of this class: they are not transmitted across generations, they carry information only in the causal sense and information transmission is not their role under any reasonable teleofunctional explanation. Biological structures such as membranes and centrosomes may appear as some sort of middle ground, but we note that (1) their primary function is not an informational one and (2) their bandwidth is extremely restricted compared to that of DNA sequence. Birds' nests \cite{SterelnyEtAl96} could be seen as an environmental analogue to these intracellular structures, whereas libraries start to push toward genes and methylation tags in their informational capacity. Developmental switches \cite{griffiths_2001} are another interesting case; these transduce environmental information but in addition to their bandwidth limitations they appear to have a more limited intergenerational transmission role. Genes may not be unique in their ability to convey information across generations --- but at the same time a transmission view makes it clear that not all components of the developmental matrix \cite{GriffithsAndGray94,griffiths_2001} enjoy parity in their informational capacities. 
	
The parity thesis typically is linked to Shannon's information theory via the claim that ``The source/channel distinction is imposed on a natural causal system by the observer.'' (Griffiths 2001, p.398) What is signal, and what is noise --- \citet{sterelny_sex_1999} take this to be merely a reflection of our interests. So must we impose our own notions of what makes an appropriate reference frame in order to single out certain components of the developmental matrix as signal and others as noise? No. We are not the ones who pick the reference frame, {\em natural selection} does. Because natural selection operates on heritable variation, it acts upon some components of the developmental matrix differently than others. For a biologist trying to understand life, the source-channel distinction is imposed not by the observer but rather by the process of natural selection from which life arose and diversified. 

\begin{figure}[hptb]
\centering
\includegraphics[width=0.85\columnwidth]{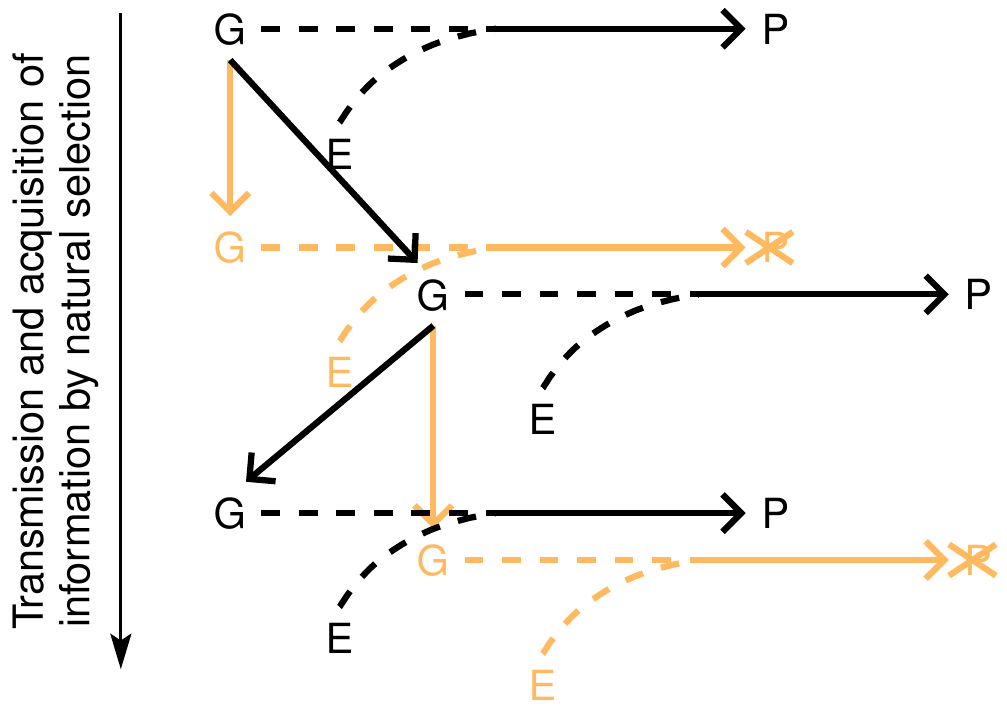}
\caption{\label{fig4}\itshape Transmission and natural selection. With the parent sending variant messages to each offspring and natural selection acting on the phenotype, information 
can accumulate in the genome.}
\end{figure}

To better understand the role of natural selection it helps to expand Figure \ref{fig2} somewhat. (For simplicity we retain our focus on the genes as transmitted elements, but one could extend this to include other heritable structures). In Figure \ref{fig2}, we highlight the fact that it is the genes, and not the environment, which is transmitted from generation to generation. In Figure \ref{fig4} we highlight the fact that {\em not all} genes are transmitted to the next generation. It is by the means of variation in the genes and selection on the phenotypes with which they are correlated that information can built up in the genome over time \cite{Felsenstein71}. 

\subsection*{Causal information versus transmitted information}

One motivation for replacing causal sense views of information with semantic-sense views is that the causal sense of information appears to cast too broad of a net. Any physical system with correlations among its components carries causal information --- but in their use of the information concept, biologists appear to mean something stronger than the notion of natural meaning that has smoke in the sky carrying information about a fire below \cite{godfrey-smith_hull_cambridge_2008}. If we substitute a transmission view of information for a semantic view, will we be driven back to this overly-broad notion of information? Not at all. Like naturalized views of semantics, the transmission notion of information rests upon function: to say that X carries information, we require that the function of X be to reduce  uncertainty on the part of a receiver. 

The failure to consider function when talking about information sometimes generates confusion among practicing biologists. After all, there are correlations everywhere in biological systems, measuring them is what we do as biologists, and we often talk about these correlations as information. This language is understandable; indeed these correlations provide {\em us} with information about biological systems. But this is merely causal sense information \cite{GodfreySmith99}. As Godfrey-Smith explains, when a systematist uses gene sequences to make inferences about population history, ``there is no more to this kind of information than correlation or `natural meaning'; the genes are not {\em trying to tell us} about their past.'' \cite{GodfreySmith99}. In other words, these correlations do not convey teleosemantic information. Nor can these correlations be considered information in the transmission sense.  

To expand upon this distinction, an example from population genetics is helpful. Voight and colleagues \citep{pritchard} developed a method for inferring positive selection at polymorphic loci in the human genome. Their key insight is that with enough sequence data from sufficiently many members of the population,  we can pick out regions of the genome that have unusually long haplotypes of low diversity. 
Such extended haplotype blocks tend to surround an allele that has recently risen in frequency due to strong selection, because there has not been enough time for recombination to break down the association between the favored allele and the genetic background in which it arose. Using this method, Voight and colleagues find strong evidence for recent selection among Europeans in the lactase gene LCT, which is important for metabolism of lactose beyond early childhood. The favored allele results from a single nucleotide change 14 kb upstream of the lactase gene
on the nearly 1 Mb haplotype. Positive selection has presumably occurred because the ability to process lactase throughout life became advantageous with the invention of animal agriculture approximately 10,000 years ago.

What does this have to do with information? There is information about the history of selection in the population-level correlations. Voight et al. found an extended haplotype length surrounding the LCT\verb-+- allele relative to that around the LCT\verb+-+ allele. But notice that we have to observe the genotypes of multiple individuals in order to determine that one allele at the LCT locus is surrounded by longer haplotype blocks than is the other.  Once we have made observations of multiple genomes, we as external observers can conclude something about the history of selection on the population. But this information is not available at the level of a single individual. A single individual cannot look at its own genome and notice a longer (or shorter) haplotype block around any given focal locus --- these haplotype blocks are defined with respect to the genotypes of others in the population. A single individual can only look at its own genome and see a sequence of base pairs. This sequence of base pairs is what is transmitted, it is what has the function of reducing uncertainty on the part of the agent who observes it\footnote{Although causal-sense information is transmitted from the population at time t to the population at time t+1 in the population frequencies of haplotypes, this is not transmission-sense information because the function of these population-level haplotype assemblages is not to reduce uncertainty on the part of future populations.}. These individual gene sequences are the entities that have an informational function in biology (though bioinformaticians have not always recognized this distinction \cite{Adami02}). The population-level statistics that geneticists use to infer history are informative but they are not information in the transmission sense. They are merely the smoke that is cast off by the fire of natural selection.

\subsection*{Coding without appeal to semantics}

The transmission sense of information allows us to separate claims about how information is transmitted from claims about what information means\footnote{The source-channel separation theorem \nocite{CoverAndThomas06} (Cover and Thomas 2006, Chapter 7, p.218) proves that in any physical communication system for error-free transmission over a noisy channel, one can entirely decouple the process of tuning the code to the nature of the specific channel from not only the semantic reference of the signal but indeed from all statistics of the message source. This follows because the theorem states that one can achieve channel capacity with separate source and channel coders --- and in this setup the source coder can always be configured so as to return output that maximizes the entropy given the symbol set.}.
Indeed, we can study how information is transmitted without having any knowledge of the ``codebook'' for how to interpret the message, or even what the information represents.

In many biological studies, we are in exactly this position. Again an example --- this time drawn from neurobiology --- is helpful. In a study of the fly visual system, de Ruyter van Steveninck and colleagues  presented flies with a moving grating as a visual stimulus, and made single-cell recordings of the spike train from the H1 visual neuron \cite{deRuyterVanSteveninckEtAl97}. This neuron is sensitive to movement, but it is not known how movement information is encoded into the spike train, nor even what aspects of movement are being represented. Nonetheless de Ruyter van Steveninck and colleagues were able to determine how much information this neuron is able to encode. The investigators exposed a fly to the stimulus, and measured the (average) entropy of the spike train. This is the so-called total entropy for the neuron's output. They then looked at what happens if you play the same stimulus back repeatedly: how much does the resultant spike train vary from previous trials? This is the so-called noise entropy. The information that the spike train carries about the stimulus, i.e., the mutual information between spike train and stimulus, is simply the difference of these two quantities. Using this approach, the authors were able to show that this single insect neuron conveys approximately 80 bits of information per second about a dynamic stimulus. Thus an individual visual neuron achieves a bit-rate that is roughly 7 times the bit rate of a skilled touch typist\footnote{Using Shannon's 1950 upper bound on bits per letter and his estimate of letters per word in the English language  \cite{Shannon50}, we can estimate the bit rate of a touch typist as $ \frac{120  \text{ words}}{\text{ minute}} \frac{1 \text{ minute}}{\text{60 seconds}} \frac{4.5 \text{ letters}}{\text{ word}} \frac{1.3 \text{ bits}}{\text {letter}} =11.7 \text{ bits/second}$. }! More importantly, the researchers were able to compare the response of this neuron to static stimuli with the response of the neuron to natural patterns of motion. They found that for natural patterns the neuron is able to attain the high bandwidth that it does by ``establishing precise temporal relations between individual action potentials and events in the sensory stimulus''. By doing so, the neuron's response to the dynamic stimulus greatly surpasses the bit rate that could be obtained if the stimulus was encoded by a simple matching of spike rate to stimulus intensity. Subsequent investigators have used related methods to show that evolved sensory systems are tuned to natural stimuli, to study the properties of neural adaptation and history dependence, and to examine temporal sensitivity --- all without knowing the way in which the signals that they study are actually encoded.

de Ruyter van Steveninck et al.\/ sum up the power of being able to study information without appeal to semantics: 	``This characterization of ... information transmission is independent of any assumptions about [or knowledge of!] which features of the stimulus are being encoded or about which features of the spike train are most important in the code''  \cite{deRuyterVanSteveninckEtAl97}

This brings us back to Shannon theory. When information theorists think about coding, they are not thinking about
semantic properties. All of the semantic properties are stuffed into the
codebook, the interface between source structure and channel structure,
 which to information theorists is as interesting as a phonebook is
to sociologists. When an information theorist says ``Tell me how data
stream A codes for message set B'', she is not asking you to read her the
codebook. She is asking you to tell her about compression, channel capacity,
distortion structure, redundancy, and so forth. That is, she wants to know how the
structure of the code reflects the statistical properties of the data source and the channel, with respect to the decision problem of
effectively packaging information for transport.

With these things in focus, we can now look at the concept of arbitrariness, what it means, and why this concept is critically important in biological coding.

\section{Information theory and arbitrariness}

In arguing that DNA is an informational molecule, Maynard Smith \cite{maynardsmith2000} appeals to Jacques Monod's concept of {\em gratuit\'e} \cite{monod71}, and a number of additional authors have further explored this thread \cite{godfrey-smith_theoretical_2000,godfreysmith_maynardcomment2000,stegmann05}. For Monod, {\em gratuit\'e} was an important component of the logical structure of his theory of gene regulation. {\em Gratuit\'e} is the notion that in principle regulatory proteins can cause any inducer or repressor to influence the expression of any region of DNA. There need be no direct chemical relation between the structure of an inducer and the nucleic acid sequence on which it operates.  Maynard Smith observes that we can see something like {\em gratuit\'e} in the structure of the genetic code as well: there is an {\em arbitrary} association between codons and the amino acids that they specify. 

Yet as they grapple with this idea of an arbitrary code, these authors confront the fact that the genetic code is not a random assignment of codons to amino acids, but rather a one-in-a-million evolved schema for associating these molecules: the genetic
code is structured so as to smooth the mutational landscape and ensure that common translational errors  generate
amino acid replacements between chemically 
similar amino acids \cite{HaigAndHurst91,freeland_genetic_1998}.
So what can these authors mean when they say that the code is arbitrary? Maynard Smith, Godfrey-Smith, and others are not suggesting  that the structure of the code is random or contingent on random historical processes as in Crick's frozen accident hypothesis. Rather, they are making a semiotic claim. Arbitrariness refers to the fact that ``[m]olecular symbols in biology are symbolic,'' as opposed to indexical or iconic \cite{maynardsmith2000}. In the case of the genetic code, this means that the association between a codon and its corresponding amino acid is not driven by the immediate steric interactions between the codon and the amino acid, but instead is mediated by an extensive tRNA structure that in principle could have coupled this codon to any other amino acid instead. 

This fact is enormously important to the function of the biological code --- not as a matter of the semiotic classifications that fascinated Charles Pierce, but rather to solve the sort of decision problem that motivated Claude Shannon. From the symbolic relation between code and product, there arise the degrees of freedom that a communication engineer requires to tune the code to the statistical properties of source and channel.  To see how this works we will visit an example from the early history of telecommunications.

For over one hundred years, Morse code was the standard protocol for telegraph and radio communication. The code transcribes the English alphabet into codewords composed of short pulses called dots and long pulses called dashes. For example, the letter E is represented by a single dot ``\verb+.+'', the letter T is represented by a single dash ``\verb+-+'', the letter Q by the quartet ``\verb+--.-+'', and the letter J by ``\verb+.---+''. At first glance, the mapping between letters and Morse codewords appears to be arbitrary.  They are certainly symbolic rather than iconic or indexical. But there is an important pattern to the way that letters are assigned codes in Morse code. 

But instead of assigning codewords sequentially, ``\verb+.+'' to A, 
``\verb+-+'' to B, ``\verb+..+'' to C, Samuel Morse exploited the degrees of
freedom available for codeword assignments to make an efficent code for fast transmission
of English sentences. He could not assign short code words to every single letter --- there simply are not enough short code words to go around. Instead, by assigning the shortest code words to the most commonly used letters, Morse created a code in which transmissions would on average be shorter than if he had used sequential codeword assignments.

Morse could not have done this with pictograms.
The leeway to associate any message with any code word provides the communications engineer
with the degrees of freedom that he needs to tune the semantic
and statistical properties of the source messages to the transmission cost and
error properties of the channel. In the absence of a full
picture of the engineer's decision problem, the code might look arbitrary. But a well-chosen code is not arbitrary at
all; it solves a decision problem for packaging.

Morse exploited the available degrees of freedom in his choice of codes; apparently natural selection has done the same in evolving a one-in-a-million genetic code. Thinking about these degrees of freedom --- along with an important thought experiment --- helped us to understand the role of coding in the transmission sense of information.  \citet{godfrey-smith_theoretical_2000} imagined a hypothetical world composed of ``protein genes'' as a way to explore the importance of the coding concept in biology. Since the idea of coding presumably refers to the fact that DNA provides an arbitrary combinatorial representation of amino acid sequence, Godfrey-Smith considers the nature of a world in which the hereditary material was neither arbitrary nor representational. He images a world of ``protein-genes'' in which there is no translation but instead a copying system in which amino acid sequences,  assisted by coupling molecules, replicate using previous amino acid sequences as templates. In that world, Godfrey-Smith argues, there is nothing that corresponds to coding, and yet stepping away from the microscope, biology functions much as before. In Godfrey-Smith's protein-sample world, there is no code, no compression, no redundancy, and information theory can merely be applied as a descriptive statistics for correlations. One could even go so far as to argue that in a protein-genes world, physical structure and not information is inherited across generations. Thus Godfrey-Smith concludes that ``Removing genetic coding from the world need not change much else, and this gives support to my claim that we should only think of coding as part of an explanation of how cells achieve the specific task of putting amino acids in the right order,'' rather than something fundamental to the logical structure of biology. 

But there is a critical difference between a world with a proper genetic code and world based upon protein-genes: only the former allows an arbitrary combinatorical mapping between templates and products. From an information-theoretic perspective, this is absolutely critical. The degrees of freedom to construct an arbitrary mapping of this sort turn the problem of code evolution from one of passing physical samples to the next generation, into a decision-theoretic problem of how to package information for transport. In the protein-genes world, the fidelity of transmission is at the mercy of the biochemical technology for copying. There are no degrees of freedom for structuring redundancy, minimizing distortion, or conducting the other optimization activities of a communications engineer. In a DNA-based code, the chemically arbitrary assignments of nucleotide triplets to amino acids via tRNAs offer the degrees of freedom to do all of the above. While the precise dynamics of code evolution remain unknown, it appears that natural selection has put these degrees of freedom to good use. 

The difference between a protein-genes world and a DNA-based world became clear by taking the perspective of a communications engineer. Throughout the paper this has been our approach. Correlations, symmetry of mutual information, the parity thesis, arbitrariness, coding --- all these these come into focus from a communications viewpoint. We see what makes the genetic code a code, and we get a new perspective on the information language that is part of the everyday working vocabulary of researchers in genetics and evolutionary biology. The transmission sense of information justifies such language as more than shallow metaphor.

%\bibliography{shannonbiology}

\end{document}